\documentclass[twocolumn,showpacs,preprintnumbers,amsmath,amssymb]{revtex4}

\usepackage{graphicx}
\usepackage{dcolumn}
\usepackage{bm}

\begin{document}

\preprint{APS/123-QED}

\title{Compressible Sub-Alfvenic MHD turbulence\\
       in Low $\beta$ Plasmas}

\author{Jungyeon Cho}
\author{A. Lazarian}%
\affiliation{%
Astronomy Dept., Univ.~of Wisconsin, Madison, WI53706, USA; 
 cho, lazarian@astro.wisc.edu 
}%


\date{\today}

\begin{abstract}
We present a model for compressible sub-Alfv\'{e}nic isothermal 
magnetohydrodynamic (MHD) turbulence
in low $\beta$ plasmas and numerically test it.
We separate MHD fluctuations into 3 distinct families - Alfv\'{e}n,
slow, and fast modes.
We find that, 
production of slow and fast modes by Alfv\'{e}nic turbulence
is suppressed.
As a result, Alfv\'{e}n modes in compressible regime exhibit scalings 
and anisotropy
similar to those in incompressible regime.
Slow modes passively mimic Alfv\'{e}n modes.
However, fast modes show isotropy and a scaling similar to 
acoustic turbulence.

\end{abstract}

\maketitle

\section{\label{sec:intro}Introduction}


Most astrophysical fluids, including stellar winds and 
the interstellar medium (ISM),
are turbulent \cite{ARS95,CLV02c}
with an embedded magnetic field that influences almost all of
their properties.
High interstellar Reynolds numbers ($Re \equiv L\delta V/\nu > 10^{8}$; 
L=the characteristic
scale or driving scale 
of the system, $\delta V$=the velocity difference over this scale, 
and $\nu$=viscosity)
ensure that.
Turbulence spans from km to
kpc scales
and holds the key to many astrophysical
processes (e.g.,
star formation, fragmentation of molecular
clouds, heat and cosmic ray transport, magnetic reconnection).
Statistics of turbulence is also essential for the CMB foreground
studies \cite{Laz02}.

Kolmogorov scalings \cite{Kol41} were the first major
advance in the theory of incompressible (non-magnetized) turbulence.
Kolmogorov theory predicts an isotropic power law energy spectrum 
($E(k)\propto k^{-5/3}$) in wave-vector space ${\bf k}$.

Attempts to describe magnetic incompressible 
turbulence statistics were made by
Iroshnikov \cite{Iro63} and Kraichnan \cite{Kra65}. Their model
of turbulence (IK theory) 
is isotropic in spite of the presence of the magnetic field and
predicts $k^{-3/2}$ power law energy spectra for both velocity and
magnetic field.
However, the assumption of isotropic energy distribution in wave-vector space
has been criticized by many researchers \cite{Mon82,SMM83}.

An ingenious model very similar in its beauty and simplicity 
to the Kolmogorov model has been proposed by 
Goldreich \& Sridhar \cite{GS95} (hereinafter GS95) 
for incompressible MHD turbulence. 
It predicts a Kolmogorov-like energy spectra 
($E(k_{\perp})\propto k_{\perp}^{-5/3}$) 
in terms of wave-vector component $k_{\perp}$ which is perpendicular to the
local direction of magnetic field.
The parallel component of the wave-vector $k_{\|} \propto k_{\perp}^{2/3}$
within the model.
Numerical simulations \cite{CV00a,MG01,CLV02a} support the GS95 model.

In this paper, we study compressible supersonic sub-Alfv\'{e}nic MHD turbulence
in low-$\beta$ plasmas.

\section{Theoretical considerations}
While the GS95 model describes incompressible
MHD turbulence well,
no accepted theory exists 
for compressible MHD turbulence.
Earlier theoretical and numerical efforts \cite{Hig84,ZM93,MGOR96}
addressed effects of
compressibility for limited parameter spaces.
In (isothermal) plasmas, 
there are 3 types of MHD waves - Alfv\'{e}n, slow, and
fast waves.
Alfv\'{e}n modes are incompressible while
slow and fast modes are compressible.
Lithwick \& Goldreich \cite{LG01} conjectured that
Alfv\'{e}n modes follow the GS95 model and slow modes passively follow
the same scalings for high $\beta$ ($\equiv P_g/P_B\approx 2a^2/V_A^2$; 
$P_g$=gas pressure,
$P_B$=magnetic pressure; $a$=sound speed; $V_A$=Alfv\'{e}n speed) regime, 
which is 
largely similar to the exactly incompressible regime.
They also mentioned that this relation can carry on
for low $\beta$ plasmas.

In the ISM $\beta$  is frequently less than unity.
For instance, it is $\sim 0.1$ or less for molecular clouds.
Therefore, we consider low $\beta$ regime in this paper.
Interstellar turbulence is traditionally thought to be 
sub-Alfv\'{e}nic ($\delta V < V_A$), although this is not a universally accepted
assumption (see e.g.~\cite{Bol02}).
If turbulence is super-Alfv\'{e}nic initially, we expect that eventually
magnetic energy should approach the equipartition level \cite{CV00g} and
the scales smaller than the energy injection scale 
should fall in the sub-Alfv\'{e}nic compressible regime.

Arguments in GS95 are suggestive that the coupling of Alfv\'{e}n 
to fast and slow modes 
will be weak. Consequently, we expect that in this regime the Alfv\'{e}n
cascade should follow the GS95 scaling.
Moreover the slow modes are likely to evolve passively (see \cite{LG01}), 
so that we expect the GS95 scaling for them as well.
However, fast modes are expected to show isotropic distribution
as their velocity does not depend on magnetic field direction.
To test those theoretical conjectures we use numerical
simulations.

\section{ Numerical Method}
To mitigate spurious oscillations near shocks, we
combine two essentially non-oscillatory (ENO) schemes.
When variables are sufficiently smooth, we use the 3rd-order
Weighted ENO scheme \cite{JW99}
without
characteristic mode decomposition. 
When opposite is true, we use the 3rd-order Convex ENO scheme \cite{LO98}.
We use a three-stage Runge-Kutta method for time integration.
We solve the ideal MHD equations in a periodic box:
\begin{eqnarray}
{\partial \rho    }/{\partial t} + \nabla \cdot (\rho {\bf v}) =0, \nonumber \\
{\partial {\bf v} }/{\partial t} + {\bf v}\cdot \nabla {\bf v} 
   +  \rho^{-1}  \nabla(a^2\rho)
   - (\nabla \times {\bf B})\times {\bf B}/4\pi \rho ={\bf f}, \nonumber \\
{\partial {\bf B}}/{\partial t} -
     \nabla \times ({\bf v} \times{\bf B}) =0,\nonumber 
\end{eqnarray}
with
    $ \nabla \cdot {\bf B}= 0$ and an isothermal equation of state.
Here $\bf{f}$ is a random large-scale driving force, 
$\rho$ is density,
${\bf v}$ is the velocity,
and ${\bf B}$ is magnetic field.
The rms velocity $\delta V$ is maintained to be approximately unity, so that
 ${\bf v}$ can be viewed as the velocity 
measured in units of the r.m.s. velocity
of the system and ${\bf B}/\sqrt{4 \pi \rho}$ 
as the Alfv\'{e}n speed in the same units.
The time $t$ is in units of the large eddy turnover time ($\sim L/\delta V$) 
and the length in units of $L$, the scale of the energy injection.
The magnetic field consists of the uniform background field and a
fluctuating field: ${\bf B}= {\bf B}_0 + {\bf b}$.

For mode coupling studies (Fig.~\ref{fig_coupling}), 
we use $144^3$ grid points and we do {\it not} drive turbulence.
We explicitly vary the Alfv\'{e}n speed of 
the background field, $V_A=B_0/\sqrt{4 \pi \rho_0}$,
and/or the sound speed. Here $\rho_0$ is the average density.
For scaling studies (Fig.~\ref{fig_scaling}), 
we drive turbulence solenoidally in Fourier space and
use $216^3$ points, $V_A=1$, $\rho_0=1$, and $a=\sqrt{0.1}$.
The average rms velocity in statistically stationary state is 
$\delta V\sim 0.7$.
Therefore, the scaling results reported here utilize 
$M_s (=\delta V/a)\sim 2.2$, $M_A (=\delta V/V_A)\sim 0.7$, 
and $\beta \sim 0.2$.

\section{Results}

\begin{figure*}[!t]
\begin{tabbing}
~\includegraphics[width=.33\textwidth]{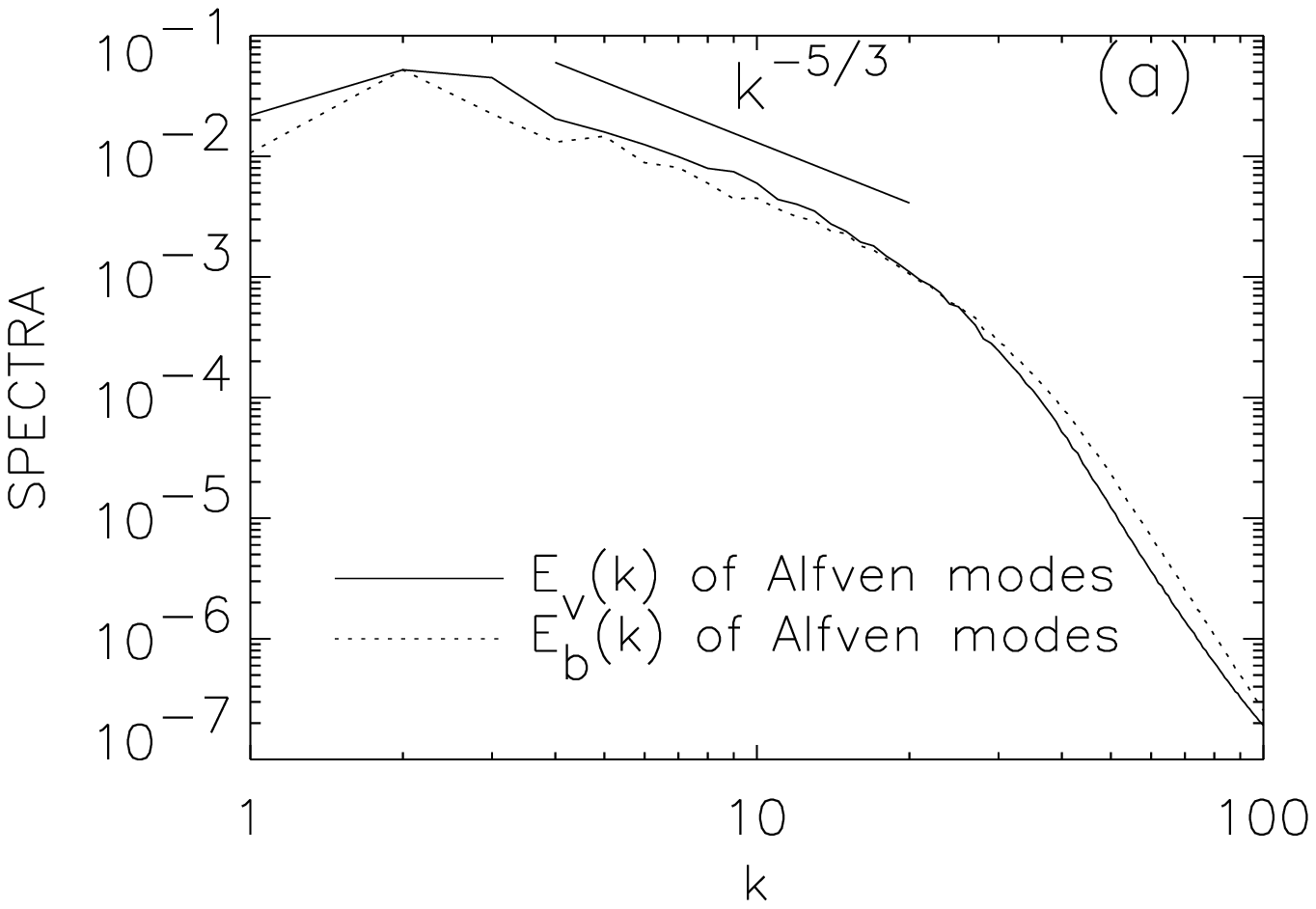}
\=
\includegraphics[width=.33\textwidth]{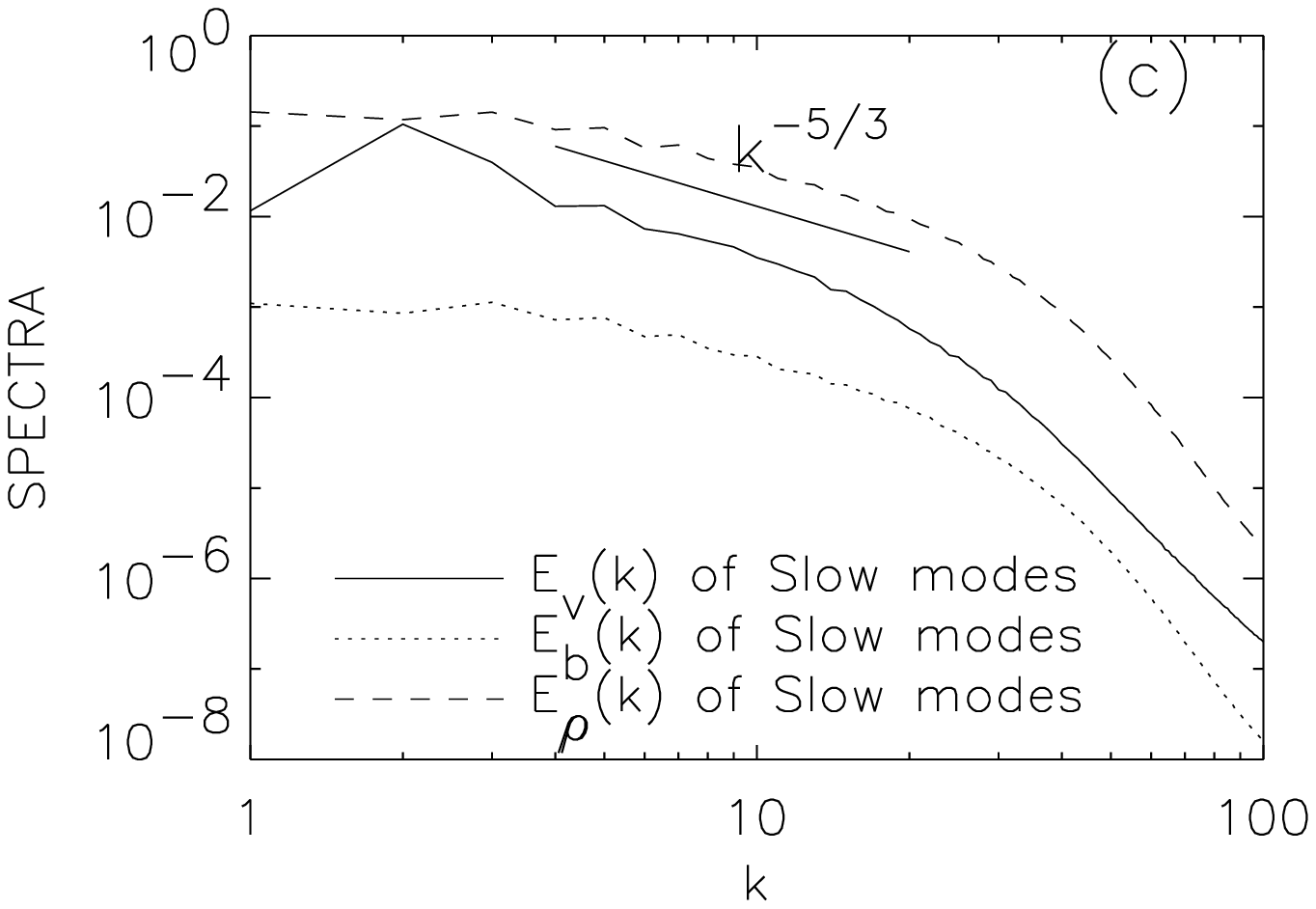}
\=
\includegraphics[width=.33\textwidth]{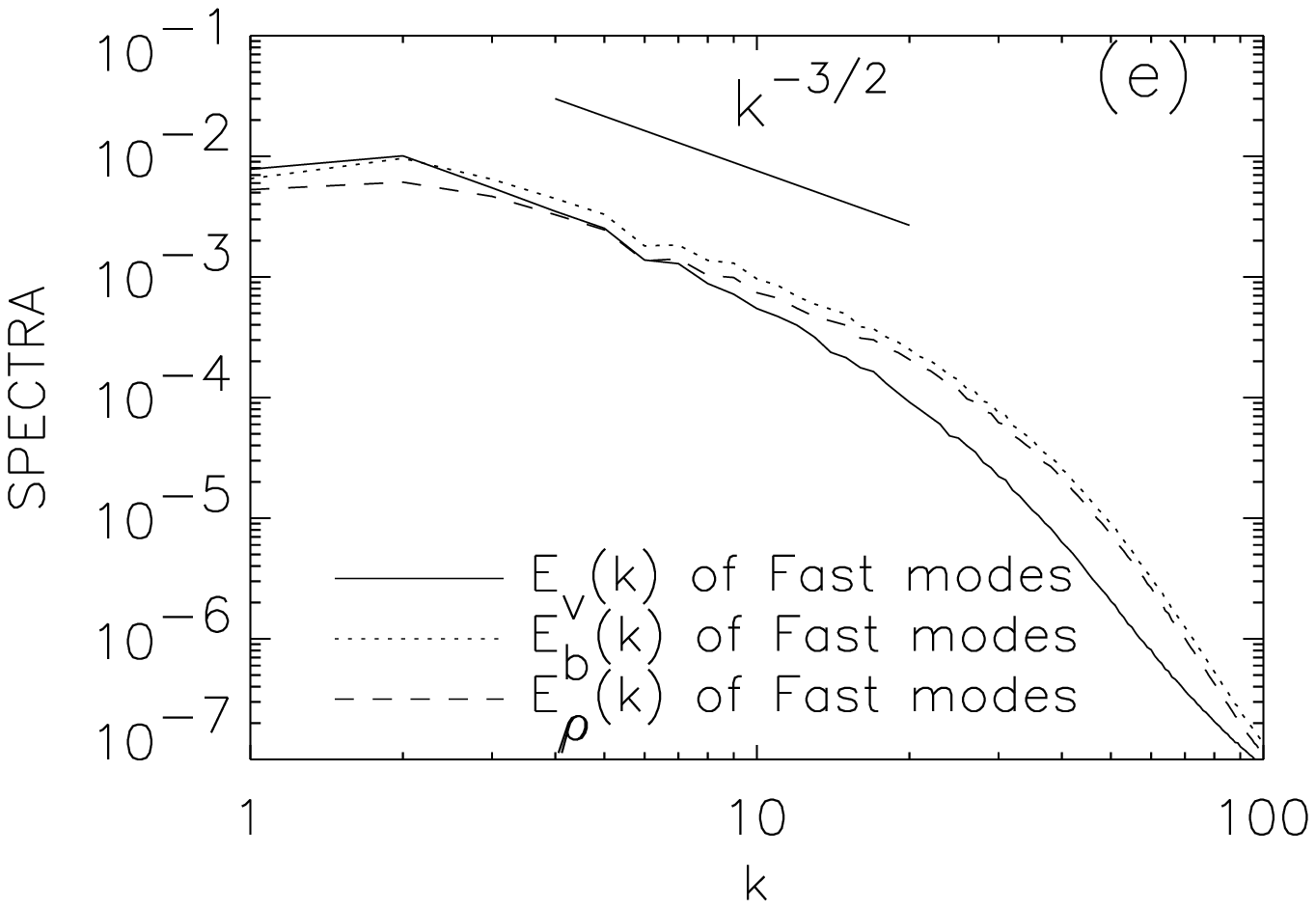}
\\  \\
\includegraphics[width=.33\textwidth]{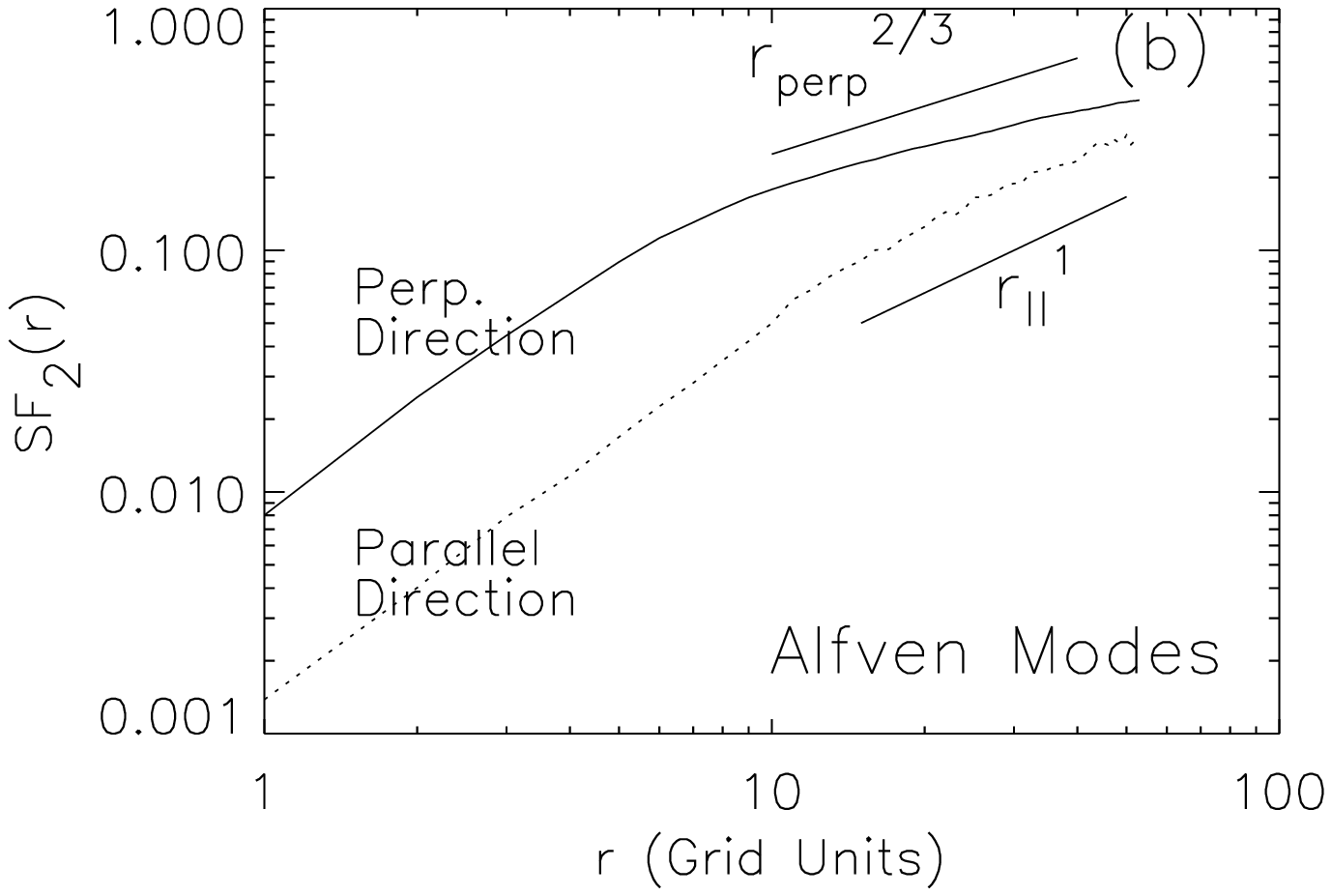}
\=
~~~~\includegraphics[width=.30\textwidth]{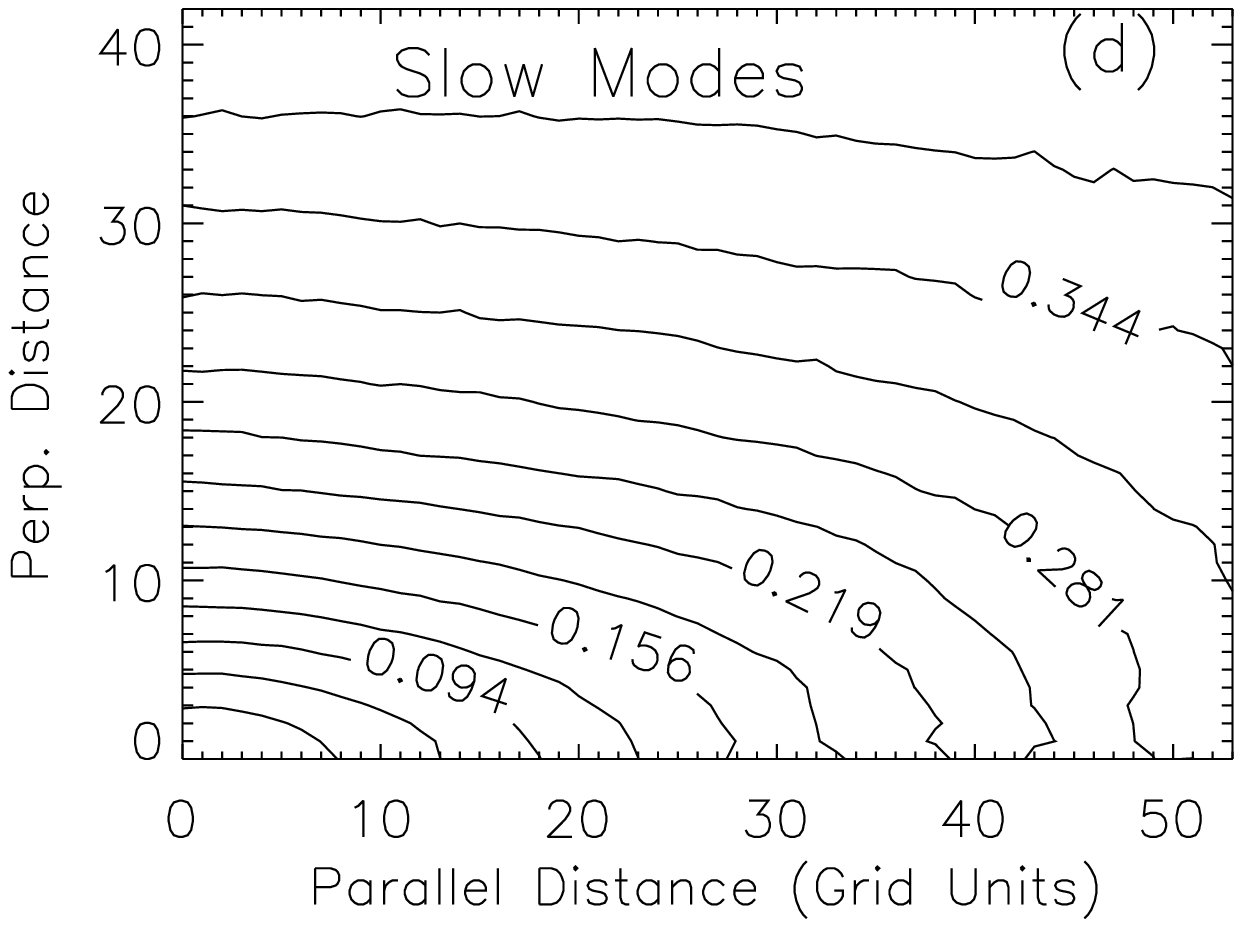}
\=
~~~~~\includegraphics[width=.30\textwidth]{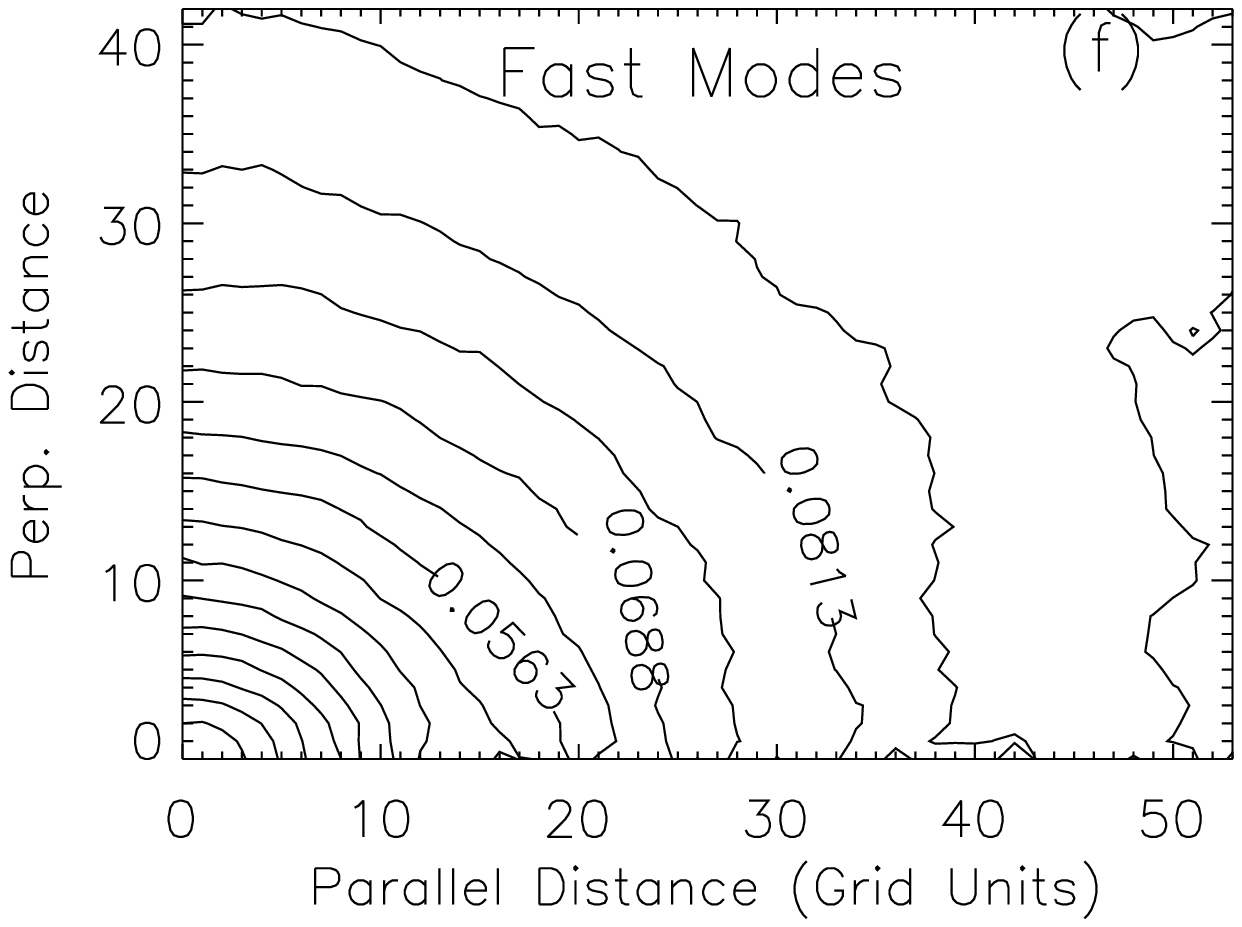}
\end{tabbing}
\caption{ Scalings relations. Results from driven turbulence with
          $M_s\sim 2.2$, $M_A\sim 0.7$, $\beta\sim 0.2$, 
          and $216^3$ grid points.
         (a) Spectra of Alfv\'en modes follow a Kolmogorov-like power law.
         (b) The second-order structure function 
             ($SF_2$) 
             for velocity of Alfv\'en modes
             shows anisotropy similar to the GS95
        ($r_{\|}\propto r_{\perp}^{2/3}$ or $k_{\|}\propto k_{\perp}^{2/3}$).
         The structure functions are measured in directions perpendicular or
             parallel to the local mean magnetic field in real space.
             We obtain real-space velocity and magnetic fields 
             by inverse Fourier transform of
             the projected fields.
         (c) Spectra of slow modes also follow a Kolmogorov-like power law.
         (d) Slow mode velocity shows anisotropy similar to the GS95.
             We obtain contours of equal $SF_2$ directly in real space
             without going through the projection method,
             assuming slow mode velocity is nearly parallel to local mean
             magnetic field in low $\beta$ plasmas.
         (e) Spectra of fast modes are compatible with
             the IK spectrum.
         (f) The magnetic $SF_2$ of 
             fast modes shows isotropy.
             We obtain real-space magnetic field 
             by inverse Fourier transform of
             the projected fast magnetic field.
             Fast mode velocity also shows isotropy.
       }
\label{fig_scaling} 
\end{figure*}

\noindent
{\bf Mode coupling of MHD waves.---}
We first describe how to separate Alfv\'{e}n, slow, and fast modes
in wave-vector (or, Fourier) space.
In general, displacement vectors (hence ${\bf v}_k$) 
of slow waves and fast waves are
\begin{eqnarray}
   \hat{\bf \xi}_s &\propto &
        k_{\|} \hat{\bf k}_{\|}+
     \frac{ 1-\sqrt{D}-{\beta}/2  }{ 1+\sqrt{D}+{\beta}/2  } 
    \left[ \frac{ k_{\|} }{ k_{\perp} }  \right]^2
     k_{\perp} \hat{\bf k}_{\perp},  \label{eq_xis}     \\
   \hat{\bf \xi}_f &\propto &
     \frac{ 1-\sqrt{D}+{\beta}/2  }{ 1+\sqrt{D}-{\beta}/2  } 
    \left[ \frac{ k_{\perp} }{ k_{\|} } \right]^2
     k_{\|} \hat{\bf k}_{\|}  +
          k_{\perp} \hat{\bf k}_{\perp},
\end{eqnarray}
where $D=(1+{\beta}/2)^2-2{\beta} \cos^2{\theta}$ and 
$\theta$ is the angle between ${\bf k}$ and ${\bf B}_0$.
In the limit of $\beta \rightarrow 0$, the displacement vectors of
the slow waves are almost parallel to ${\bf k}_{\|}$ ($||~{\bf B}_0$) and
those of fast modes are almost parallel to ${\bf k}_{\perp}$ 
($\perp {\bf B}_0$).
We can obtain slow and fast velocity by projecting velocity component 
${\bf v}_{\bf k}$ into $\hat{\bf \xi}_s$ and $\hat{\bf \xi}_f$, respectively.
We can obtain velocity and magnetic field due to Alfv\'{e}n modes
in the same
way as in incompressible case (see \cite{MG01}): 
$\hat{\bf \xi}_A = \hat{\bf k}_{\|} \times \hat{\bf k}_{\perp}$.
{}To separate slow and fast magnetic modes, 
we assume the linearized continuity equation
($\omega \rho_k = \rho_0 {\bf k} \cdot  {\bf v}_k$) and
the induction equation
($\omega {\bf b}_k = {\bf k} \times ({\bf B}_0 \times {\bf v}_k)$)
are {\it statistically} true.
{}From these, we get Fourier components of density
and {\it non-Alfv\'{e}nic} magnetic field:
\begin{eqnarray}
\rho_k 
       &=&(\rho_0 \Delta v_{k,s}/c_s) \hat{\bf k}\cdot \hat{\bf \xi}_s
      +(\rho_0 \Delta v_{k,f}/c_f) \hat{\bf k}\cdot \hat{\bf \xi}_f
        \nonumber \\
       &\equiv&\rho_{k,s}+\rho_{k,f},   \label{eq_rho}   \\
b_k    &=&
       (B_0 \Delta v_{k,s}/c_s) |\hat{\bf B}_0\times \hat{\bf \xi}_s|
      +(B_0 \Delta v_{k,f}/c_f) |\hat{\bf B}_0\times \hat{\bf \xi}_f|
     \nonumber \\
    &\equiv&b_{k,s}+b_{k,f}   \label{eq_b1}  \\
    &=& \rho_{k,s} (B_0/\rho_0)
      (|\hat{\bf B}_0\times \hat{\bf \xi}_s|/\hat{\bf k}\cdot \hat{\bf \xi}_s)
                                     \nonumber   \\
      &+& \rho_{k,f} (B_0/\rho_0)
      (|\hat{\bf B}_0\times \hat{\bf \xi}_f|/\hat{\bf k}\cdot \hat{\bf \xi}_f),
             \label{eq_b2}
\end{eqnarray}
where $\Delta v_k \propto v_k^+-v_k^-$ (superscripts `+' and `-'
represent opposite directions of wave propagation) 
and subscripts `s' and `f' stand for `slow' and
`fast' modes, respectively. 
{}From equations (\ref{eq_rho}), (\ref{eq_b1}), and (\ref{eq_b2}),
we can obtain $\rho_{k,s}$, $\rho_{k,f}$, $b_{k,s}$, and $b_{k,f}$
 in Fourier space.
We obtain energy spectra (Fig.~\ref{fig_scaling}a,c,e) 
using this projection method done in
Fourier space.
When we calculate 
structure functions (Fig.~\ref{fig_scaling}b,f)
we first obtain the Fourier components using the projection and, then, 
we obtain the real space values by performing inverse Fourier transform of 
the projected components.
However, we use a different method for the structure function of 
slow mode velocity (see Fig.~\ref{fig_scaling}d).

The dispersion relation of Alfv\'{e}n modes and those of slow and fast modes in
$\beta \rightarrow 0$ limit are
$ \omega = V_A k_{\|},~\omega = a k_{\|}$, and $\omega = V_A k$, respectively.
Alfv\'{e}n modes are not susceptible to collisionless damping.
Therefore, we mainly consider transfer of energy from  
Alfv\'{e}n modes to the compressible MHD ones (i.e.
slow and fast).

\begin{figure}[t]
\includegraphics[width=.50\columnwidth]{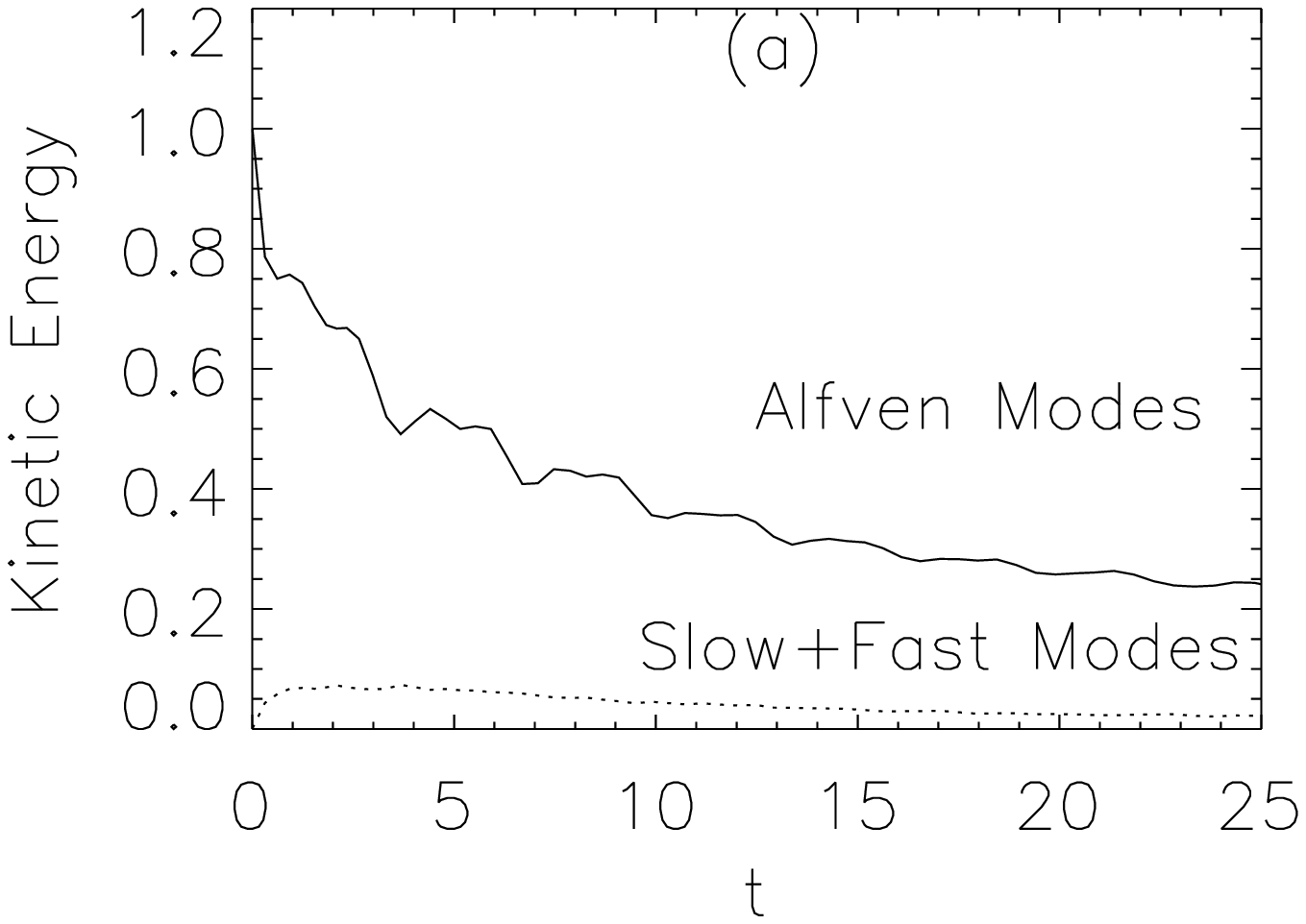}
\hfill
\includegraphics[width=.48\columnwidth]{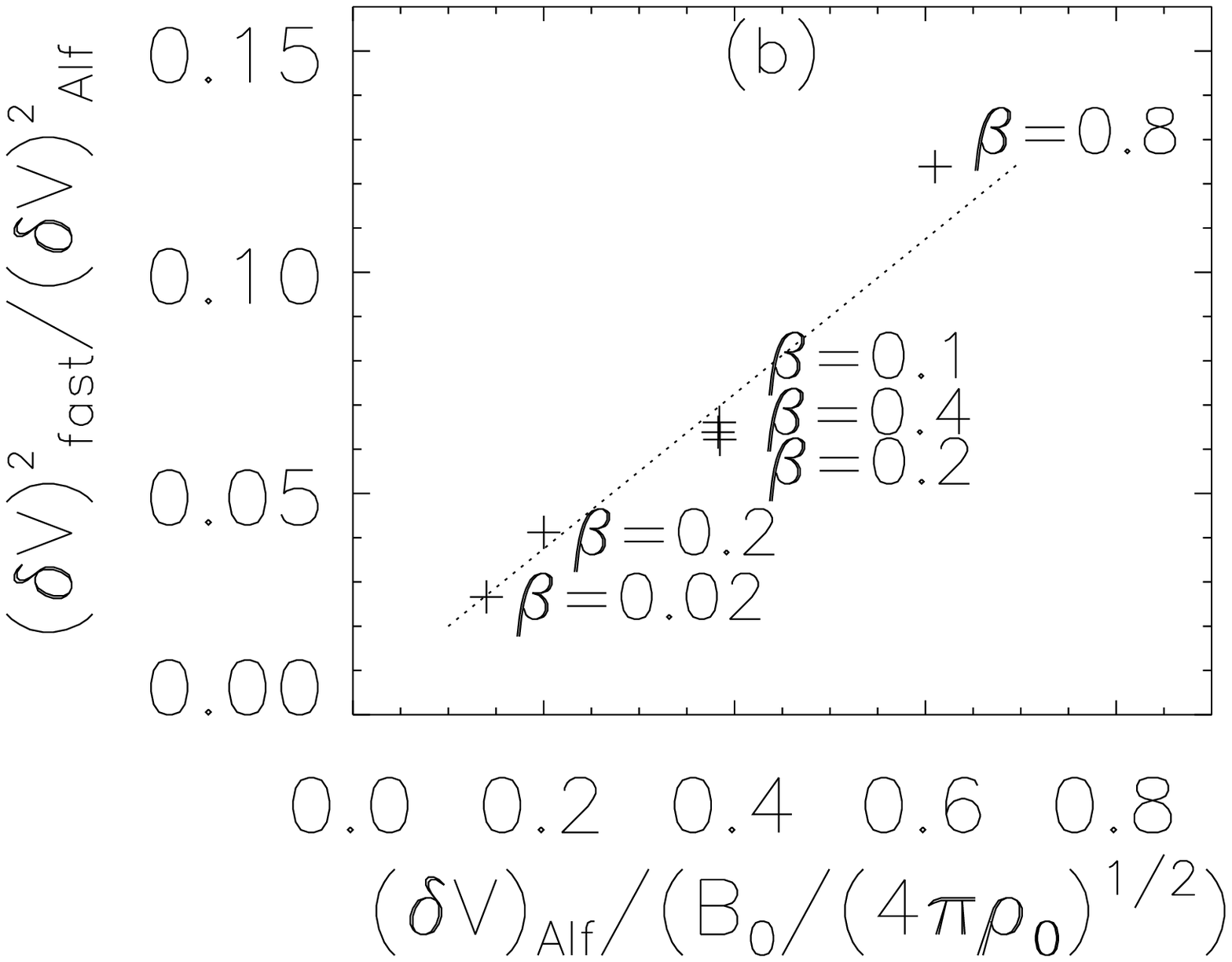}
\caption{
         ({\it a}) Decay of Alfv\'{e}nic turbulence.
         The generation of fast and slow waves is not efficient.
         Initially, $\beta \sim 0.2$ and $B_0/\sqrt{4 \pi \rho_0}=1$.
         ({\it b}) The ratio of $(\delta V)_f^2$ to
          $(\delta V)_A^2$. 
          The ratio is measured at $t \sim 3$ for all simulations.
          The ratio strongly depends on $B_0$, but only weakly on
          (initial) $\beta$.
          The initial Mach numbers span $1-4.5$.
       }
\label{fig_coupling}
\end{figure}

To check the strength of the coupling,
we first perform
a forced supersonic sub-Alfv\'{e}nic
MHD simulation with $B_0/\sqrt{4 \pi \rho_0}=1$.
Using the same data cube obtained from this simulation, we perform
several decaying MHD simulations.
We go through
the following procedures before we let the turbulence decay.
We first remove slow and fast modes in Fourier space and
retain only Alfv\'{e}n modes.
We also change the value of ${\bf B}_0$ preserving its original
direction.
We use the same constant initial density $\rho_0$ for all simulations.
We assign a new constant initial gas pressure $P_g$
 \footnote{
      The changes of both $B_0$ and $P_g$ preserve the 
      Alfv\'{e}n character of perturbations.
      In Fourier space,
      the mean magnetic field (${\bf B}_0$)
      is the amplitude of ${\bf k}={\bf 0}$ component.
      Alfv\'{e}n components in Fourier space are
      for ${\bf k} \neq {\bf 0}$ and their directions are 
      parallel/anti-parallel to
      $\hat{\bf \xi}_A$ (= $\hat{\bf B}_{0} \times \hat{\bf k}_{\perp}$).
      The direction of $\hat{\bf \xi}_A$ does not depend on
      the magnitude of $B_0$ or $P_g$.
}.
Note that $\beta=P_g/(B_0/8\pi)^2$.
After doing all these procedures, we let the turbulence decay.
We repeat the above procedures for different values of $B_0$ and $P_g$.
{}Fig.~\ref{fig_coupling}a shows time evolution of kinetic energy
of a simulation.
The solid line represents the kinetic energy of Alfv\'{e}n  modes.
It is clear that Alfv\'{e}n waves do not efficiently generate slow 
and fast modes.
Therefore we expect that Alfv\'{e}n modes follow the same scaling
relation as in incompressible case.
{}Fig.~\ref{fig_coupling}b shows that 
the following relation fits the data well:
\begin{equation}
  { (\delta V)_{f}^2 }/{ (\delta V)_{A}^2 }
 \propto { (\delta V)_{A} }/{ B_0 }, \label{vfscale}
\end{equation}
which means the coupling gets weaker as 
$B_0$ increases \footnote{ 
      The wandering of large scale (i.e.~energy injection scale) magnetic field
      may interfere with our projection method.
      However, we first note that this effect is no larger than 
      $\sim(\delta V)_{A}^2/ B_0^2$ \cite{CV00a,MG01}.
      Second, 
      the generated fast modes show isotropy similar to 
      Fig.~\ref{fig_scaling}f.
      Alfv\'{e}n modes are anisotropic.
      Therefore, it is not likely that the measured 
      isotropic mode is a numerical artifact.
}.
Note that $(\delta V)_{A}$ and $\rho_0$ are constants.
This marginal coupling
agrees well with a claim in GS95, incompressible simulations \cite{MG01},
and earlier studies where the
velocity was decomposed into a compressible component 
and a solenoidal component \cite{MGOR96,BNP01}.

\label{sec_scaling}
\vspace{0.3cm}
\noindent 
{\bf Alfv\'{e}n Modes.---} 
{}Fig.~\ref{fig_scaling}a shows that the spectra of Alfv\'{e}n waves follow
a Kolmogorov spectrum:
\begin{equation}
 \mbox{\it Alfv\'{e}n Waves:~~~~~}  E^{A}(k) \propto k_{\perp}^{-5/3}.
\end{equation}
In Fig.~\ref{fig_scaling}b, 
we plot the second-order structure function for velocity
($SF_2({\bf r})=<{\bf v}({\bf x}+{\bf r}) - 
                 {\bf v}({\bf x})>_{avg.~over~{\bf x}}$)
obtained in local coordinate systems in which the parallel axis is aligned
with the local mean field (see \cite{CV00a,CLV02a,MG01}).
The $SF_2$
 along the axis perpendicular to the local mean magnetic field
follows a scaling compatible with $r^{2/3}$.
The $SF_2$ along the axis parallel  to the local mean field follows
steeper $r^{1}$ scaling.
The results are compatible with the GS95 model 
($r_{\|}\propto r_{\perp}^{2/3}$, or $k_{\|}\propto k_{\perp}^{2/3}$).

\vspace{0.3cm}
\noindent
{\bf Slow waves.---}
The incompressible limit of slow waves is pseudo-Alfv\'{e}n waves.
Goldreich \& Sridhar \cite{GS97} argued that
the pseudo-Alfv\'{e}n waves are slaved to the shear-Alfv\'{e}n 
(i.e.~ordinary Alfv\'{e}n)
waves, which
means that
pseudo-Alfv\'{e}n modes do not cascade energy for themselves
(see also \cite{LG01}).
We confirm that similar arguments are applicable to slow waves
in low $\beta$ plasmas. 
Energy spectra in 
Fig.~\ref{fig_scaling}c  are consistent with:
\begin{equation}
 \mbox{\it Slow Modes:~~~~~}  E^{s}(k) \propto k_{\perp}^{-5/3}.
\end{equation}
In Fig.~\ref{fig_scaling}d,
contours of equal second-order 
structure function ($SF_2$), 
representing eddy shapes,
show scale-dependent isotropy: smaller eddies are more elongated.
The results are compatible with the GS95 model 
($k_{\|}\propto k_{\perp}^{2/3}$, or $r_{\|}\propto r_{\perp}^{2/3}$, where
$r_{\|}$ and $r_{\perp}$ are the semi-major axis and semi-minor axis of eddies,
respectively \cite{CV00a}).

{}From the linearized continuity equation and the induction equation, 
we can show that 
density fluctuations are dominated by
slow waves and only a small amount of magnetic field is produced by the
slow waves in low $\beta$ plasmas:
$
 ( { \delta \rho }/{ \rho })_s = { (\delta V)_s }/{ a }\sim  
             M_s,  \label{dro_ro}
$
 and  $(\delta B)_s \rightarrow 0$, as
$\beta \rightarrow 0$. 
Here $M_s$ is the sonic Mach number.
When $M_s \gg 1$, the above relation for density fluctuation
may not give a good approximation.

\vspace{0.3cm}
\noindent
{\bf Fast waves.---}
{}Fig.~\ref{fig_scaling}f shows fast modes are isotropic.
{}The resonance conditions for the interacting fast waves are
$ 
\omega_1 + \omega_2 = \omega_3 \mbox{~~and~~}
  {\bf k}_1 + {\bf k}_2 = {\bf k}_3.
$ 
Since $ \omega \propto k$ for the fast modes,
the resonance conditions can be met only when
all three ${\bf k}$ vectors are collinear.
This means that the direction of energy cascade is 
{\it radial} in Fourier space.
This is very similar to acoustic turbulence, turbulence caused by interacting
sound waves \cite{Zak67,Zak70,Lvo00}.
Zakharov \& Sagdeev \cite{Zak70} found
$E(k)\propto k^{-3/2}$.
However, there is debate about
the exact scaling of acoustic turbulence.
Here we cautiously claim that our numerical results are compatible
with the Zakharov \& Sagdeev scaling:
\begin{equation}
\mbox{\it Fast Modes:~~~~~} E^f(k) \sim  k^{-3/2}.
\end{equation}

Non-Alfv\'{e}nic magnetic field perturbations are mostly affected by
{}fast modes when $\beta$ is small:
$
   (\delta B)_f
   \sim  (\delta V)_f,
$ 
which is larger than $(\delta B)_s\approx 0$.

Turbulent cascade of fast modes is expected to be slow
and in the absence of collisionless damping
they are expected to persist in turbulent media
over longer timespans than Alfv\'{e}n or
slow modes.
This effect is difficult to observe within
numerical simulations where $\Delta B \sim B_0$.

\section{Conclusion}
We found that, in the isothermal supersonic sub-Alfv\'{e}nic low-$\beta$ 
plasmas, 
the following scalings are valid:
\begin{eqnarray*}
   &\mbox{1. Alfv\'{e}n:~} & E^A(k)  \propto k^{-5/3}, 
                        ~~~k_{\|} \propto k_{\perp}^{2/3},  \\
   &\mbox{2. Slow:~~~} & E^s(k)  \propto k^{-5/3}, 
                        ~~~k_{\|} \propto k_{\perp}^{2/3},  \\
  &\mbox{3. Fast:~~~}  & E^f(k)  \propto k^{-3/2}, 
                        ~\mbox{isotropic energy spectra}.  
\end{eqnarray*}

\noindent {\bf Acknowledgments} 
We thank Peter Goldreich for many useful suggestions,
his encouragement, and refereeing this paper.
We acknowledge the support of NSF Grant AST-0125544.
This work was partially supported by NCSA
under AST010011N and
utilized the NCSA Origin2000.


\begin{thebibliography}{8.}
\addcontentsline{toc}{section}{References}

\bibitem{ARS95} J.~W.~Armstrong, B.~J.~Rickett, \& S.~R. Spangler,
                Astrophys.~J. \textbf{443}, 209 (1995)


\bibitem{BNP01} S.~Boldyrev, A.~Nordlund, \& P.~Padoan,
                astro-ph/0111345 (2001)
\bibitem{Bol02} S.~Boldyrev, Astrophys.~J. \textbf{569}, 841 (2002)

\bibitem{CV00g} J.~Cho \& E.~T.~Vishniac,
                Astrophys.~J. \textbf{538}, 217 (2000)
\bibitem{CV00a} J.~Cho \& E.~T.~Vishniac: 
                Astrophys.~J. \textbf{539}, 273 (2000)

\bibitem{CLV02a} J.~Cho, A.~Lazarian, \& E.~T.~Vishniac, 
                Astrophys.~J. \textbf{564}, 291 (2002) 
\bibitem{CLV02c} J.~Cho, A.~Lazarian, \& E.~T.~Vishniac,
                in {\it Simulations of magnetohydrodynamic turbulence 
                     in astrophysics}, 
                eds. by T. Passot \& E.
                Falgarone (Springer Lecture Notes in Physics; 2002), submitted



\bibitem{GS95} P.~Goldreich \& H.~Sridhar,
                Astrophys.~J. \textbf{438}, 763 (1995) (GS95)
\bibitem{GS97} P.~Goldreich \& H.~Sridhar,
                Astrophys.~J. \textbf{485}, 680 (1997) 

\bibitem{Hig84} J.~C.~Higdon, Astrophys.~J. \textbf{285}, 109 (1984)

\bibitem{Iro63} P.~Iroshnikov, Astron.~Zh. \textbf{40}, 742 (1963)
               (English: Sov.~Astron. \textbf{7}, 566 (1964))
\bibitem{JW99} G.~Jiang \& C.~Wu,
                J. Comp. Phys. \textbf{150}, 561 (1999)
\bibitem{Kol41} A.~Kolmogorov,
                Dokl.~Akad.~Nauk SSSR \textbf{31}, 538 (1941)
\bibitem{Kra65} R.~Kraichnan,
                Phys.~Fluids \textbf{8}, 1385 (1965)



\bibitem{Laz02} A.~Lazarian \& S.~Prunet,
                in {\it Astrophysical Polarized Backgrounds}, 
              eds. by S. Cecchini et al. 
                (AIP Conf.~Proc.~Vol.~609, 2002) p32 (astro-ph/0111214)
\bibitem{LG01} Y.~Lithwick \& P.~Goldreich, 
               Astrophys.~J. \textbf{562}, 279 (2001)

\bibitem{LO98} X.~Liu \& S.~Osher, J.~Comp. Phys. \textbf{141}, 1 (1998)

\bibitem{Lvo00} V.~S.~L'vov, Y.~V.~L'vov, \& A.~Pomyalov, Phys.~Rev.~E, 61, 
                2586, (2000)

\bibitem{MG01} J.~Maron \& P.~Goldreich, Astrophys.~J. \textbf{554}, 1175 
                                                                     (2001)

\bibitem{MGOR96} W.~M.~Matthaeus, S.~Ghosh, S.~Oughton, \& D.~A.~Roberts,
                J.~Geophys.~Res. \textbf{101}, 7619 (1996)




\bibitem{Mon82} D.~C.~Montgomery,
                Physica Scripta \textbf{T2/1}, 83 (1982)



\bibitem{SMM83} J.~V.~Shebalin, W.~H.~Matthaeus, \& D.~C.~Montgomery,
                J.~Plasma Phys. \textbf{29}, 525 (1983)


\bibitem{Zak67} V.~E.~Zakharov, Sov. Phys. JETP, 24, 455 (1967)
\bibitem{Zak70} V.~E.~Zakharov \& A.~Sagdeev, Sov.~Phys.~Dokl. 15, 439 (1970)
\bibitem{ZM93} G.~P.~Zank \& W.~H.~Matthaeus,
                Phys.~Fluids A \textbf{5}(1), 257 (1993)


\end{thebibliography}
\end{document}